\documentclass{svproc}

\usepackage[utf8]{inputenc}
\usepackage[english]{babel}
\usepackage{bbm}
\usepackage{nicefrac}
\usepackage{slashed}
\usepackage{pstricks}
\usepackage{graphicx}
\usepackage{hyperref}
\usepackage{leftidx}
\usepackage{environ}
\usepackage{mathtools}
\usepackage{xspace}
\usepackage{suffix}
\usepackage{url}
\usepackage{slashed}

\newcommand{\ie}{\textit{i.e.}\xspace}
\newcommand{\eg}{\textit{e.g.}\xspace}

\newcommand{\apriori}{\textit{a priori}\xspace}
\WithSuffix\newcommand\apriori*{\textit{a-priori}\xspace}

\newcommand{\via}{\textit{via}\xspace}

\newcommand{\mathspace}{\ \ }
\newcommand{\mathtext}[1]{\mathspace\text{#1}\mathspace}

\newcommand{\fm}{\ensuremath{\mathrm{fm}}}

\newcommand{\veck}{\mathbf{k}}

\newcommand{\vecu}{\mathbf{u}}
\newcommand{\vecv}{\mathbf{v}}

\newcommand{\ii}{\mathrm{i}}

\newcommand{\vcalD}{\boldsymbol{\cal D}}

\newcommand{\OO}{\mathcal{O}}

\newcommand{\bra}[1]{\langle #1|}

\newcommand{\ket}[1]{|#1\rangle}

\newcommand{\braket}[2]{\langle #1|#2\rangle}

\newcommand{\mbraket}[3]{\langle #1|#2|#3\rangle}

\newcommand{\couple}[3]{\left({#1}{#2}\right){#3}}

\newcommand{\MN}{M_N}

\newcommand{\Mpi}{M_\pi}

\newcommand{\ThreeSOne}{\ensuremath{{}^3S_1}\xspace}
\newcommand{\OneSNot}{\ensuremath{{}^1S_0}\xspace}

\newcommand{\TwoH}{\ensuremath{{}^2\mathrm{H}}\xspace}

\newcommand{\Triton}{\ensuremath{{}^3\mathrm{H}}\xspace}
\newcommand{\ThreeH}{\Triton}
\newcommand{\ThreeHe}{\ensuremath{{}^3\mathrm{He}}\xspace}
\newcommand{\FourHe}{\ensuremath{{}^4\mathrm{He}}\xspace}

\newcommand{\SixLi}{\ensuremath{{}^6\mathrm{Li}}\xspace}

\newcommand*\rvec[1]%
{\ensuremath{\overset{\smash{\raisebox{-1.5pt}{\tiny$\rightarrow$}}}{#1}}}
\newcommand*\lvec[1]%
{\ensuremath{\overset{\smash{\raisebox{-1.5pt}{\tiny$\leftarrow$}}}{#1}}}

\newcommand{\MeV}{\ensuremath{\mathrm{MeV}}}

\newcommand{\sti}{\mathbf{i}}

\newcommand{\LO}{\text{LO}\xspace}
\newcommand{\NLO}{\text{NLO}\xspace}
\newcommand{\NNLO}{\text{N$^2$LO}\xspace}

\NewEnviron{subalign}[1][]{%
\begin{subequations}\begin{align}
  \BODY
\end{align}\label{#1}\end{subequations}
}

\NewEnviron{spliteq}{%
\begin{equation}\begin{split}
  \BODY
\end{split}\end{equation}
}

\begin{document}

\mainmatter

\title{The unitarity expansion for light nuclei}
\titlerunning{The unitarity expansion for light nuclei}

\author{Sebastian König\inst{1}}
\authorrunning{Sebastian König}

\institute{%
Institut für Kernphysik,
Technische Universität Darmstadt,
64289 Darmstadt, Germany
\and:q
ExtreMe Matter Institute EMMI,
GSI Helmholtzzentrum für Schwerionenforschung GmbH,
64291 Darmstadt, Germany\\
\email{sekoenig@theorie.ikp.physik.tu-darmstadt.de}
}

\maketitle

\begin{abstract}
I is argued here that (at least light) nuclei may reside in a sweet spot: bound
weakly enough to be insensitive to the details of the interaction, but dense
enough to be insensitive to the exact values of the large two-body scattering
lengths as well.
In this scenario, a systematic expansion of nuclear observables around the
unitarity limit converges.
In particular, in this scheme the nuclear force is constructed such that the 
gross features of states in the nuclear chart are determined by a very simple 
leading-order interaction, whereas---much like the fine structure of atomic
spectra---observables are moved to their physical values by small
\emph{perturbative} corrections.
Explicit evidence in favor of this conjecture is shown for the binding energies
of three and four nucleons.
\end{abstract}

\section{Introduction}
\label{sec:Introduction}

Ever since the effective range expansion (ERE) was developed as a theory to 
parameterize the low-energy two-nucleon 
system~\cite{Schwinger:1947xx,Barker:1949zz,Chew:1949zz,Bethe:1949yr} it has 
been known that the nucleon-nucleon ($N\!N$) scattering lengths, $a_t\simeq 
5.4~\fm$ and $a_s\simeq {-}23.7~\fm$ in the \ThreeSOne and \OneSNot channels, 
respectively, are large compared to the typical range of the nuclear 
interaction, $R\sim \Mpi^{-1} \simeq 1.4~\fm$, set by the inverse pion mass.
Considering quantum chromodynamics (QCD) as the underlying theory of the strong
interaction, this particular feature of the low-energy two-nucleon ($2N$) system
can be understood as an accidental ``fine tuning'' of the QCD
parameters~\cite{Beane:2001bc,Beane:2002vs,Epelbaum:2002gb,Beane:2002xf,%
Braaten:2003eu} (the quark masses) to be close to a critical point where the 
scattering lengths are infinite, the so-called ``unitarity (or unitary) limit.''

This curiosity of nature has profound consequences for the theoretical 
description of few-nucleon systems at low energies, placing them in the same 
universality class as other systems governed by large scattering lengths, such 
as cold atomic gases, where the scattering length can be tuned \via Feshbach 
resonances~\cite{Chin:2010xx}, or certain mesons which can be
interpreted as hadronic molecules~\cite{Braaten:2003he}.
Most notably, the triton is understood to be the single remaining bound state 
out of an infinite tower of Efimov states~\cite{Efimov:1970zz} that exists in
the exact unitarity limit~\cite{Bedaque:1998kg,Bedaque:1998km,Bedaque:1999ve}.
Recently, it was shown in a model-independent way that a virtual state in the
three-nucleon ($3N$) system, known to exist for a long
time~\cite{vanOers:1967lny,Girard:1979zza}, is as an S-matrix pole that would be
an excited Efimov state if nature were just a bit closer to the unitarity
limit~\cite{Rupak:2018gnc}, confirming a relation previously observed in a
separable potential model~\cite{Adhikari:1982zzb}.

Following Ref.~\cite{Konig:2016utl} it is argued here that
nature is indeed close enough to unitarity such that it is possible to 
quantitatively describe the spectra of---at least light, and possible also 
heavier---nuclei by a perturbative expansion around the limit of infinite 
two-body scattering lengths.
At leading order (LO) this yields an interaction which is parameter free in the
$2N$ sector and determined by a single three-body parameter, adjusted to keep
the triton binding energy fixed at its experimental value.
This remarkably simple theory is shown to capture the gross features of nuclei
up to \FourHe, while corrections such as the actual finite values of the $2N$
scattering lengths and electromagnetic effects are accounted for in
perturbation theory.

Quantitatively, this ``unitarity expansion'' is constructed as a variant of 
pionless effective field theory (pionless EFT).
This theory, most recently reviewed in Ref.~\cite{Hammer:2018xx}, describes 
low-energy nuclear systems in a model-independent way, guided only by the 
symmetries of QCD and the universal physics of systems governed by a large 
scattering length.
As such, it is ideally suited to set up the unitarity expansion with a minimal 
set of assumptions.
An important aspect of each EFT is the organizational principle called ``power 
counting,'' which attributes the various terms to different orders in a 
systematic expansion.
In the standard pionless theory, the expansion parameter is given by a typical 
low-momentum scale $Q$ divided by the high scale $R^{{-}1}\sim\Mpi$.
The unitarity expansion is obtained by assuming that 
$Q \sim Q_A = \sqrt{2M_NB_A/A}$, placing nuclei in a ``sweet spot''
$1/a_{s,t} < Q_A < 1/R$, where a combined expansion in $Q_AR$ and
$1/(Q_Aa_{s,t})$ converges.

In the following, the formalism is discussed in more detail by describing
its application to calculate systems of up to four nucleons.
Readers not interested in more technical details are invited to skip ahead to 
Sec.~\ref{sec:Results}, which presents the main results and provides a broader 
perspective that places the unitarity expansion in line with other recent 
results suggesting a fascinating simplification of nuclear physics.

\section{Formalism}
\label{sec:Formalism}

Following the notation of Refs.~\cite{Konig:2015aka,Konig:2016utl}, pionless EFT
is defined in terms of a Lagrange density
\begin{equation}
 \mathcal{L} = N^\dagger\left(\ii {\cal D}_0+\frac{\vcalD^2}{2\MN}\right)N
 \null + \sum\nolimits_{\sti}C_{0,\sti}
 \left(N^T P_{\sti} N\right)^\dagger \left(N^T P_{\sti} N\right)
 + D_0 \left(N^\dagger N\right)^3 +\cdots \,,
\label{eq:L-Nd}
\end{equation}
involving nonrelativistic nucleon isospin doublets $N=(p\;n)^T$ as well as
photon fields $A_\mu$ which are coupled to the nucleons \via the covariant 
derivative $\mathcal{D}_\mu = \partial_\mu + \ii eA_\mu (1+\tau_3)/2$, where $e$ 
is the proton charge and $\tau_a$ is used to label isospin Pauli matrices.
Besides these electromagnetic interactions, of which only the static Coulomb
potential is relevant to the order considered here, the theory involves only
contact (zero range) interactions proportional to ``low-energy constants''
(LECs), such as the $C_{0,\sti}$, $D_{0}$ shown in Eq.~\eqref{eq:L-Nd}, plus
other contributions (involving an increasing number of derivatives acting on the 
nucleon fields) contained in the ellipses.
The $P_{\sti}$ denote projectors onto the $N\!N$ $S$ waves, $\sti = 
\OneSNot,\ThreeSOne$, corresponding to the short-hand labels used above for the 
singlet and triplet scattering lengths.

Leading-order (LO) terms are summed up to all orders in a nonperturbative
treatment to which higher-order corrections are added in perturbation theory.
This procedure implies that the LECs of all operators are split into different
orders, \eg, $C_{0,\sti} = C_{0,\sti}^{(0)} + C_{0,\sti}^{(1)} + \cdots$.
Typically, only the leading term in this expansion introduces a new parameter
whereas the higher-order contributions are merely used to maintain lower-order 
renormalization conditions as additional corrections are included.
The unitarity expansion departs from this scheme by moving the introduction of 
two-body parameters from $C_{0,\sti}^{(0)}$ to $C_{0,\sti}^{(1)}$.

\begin{figure}[htbp]
\centering
\includegraphics[clip,width=0.9\columnwidth]{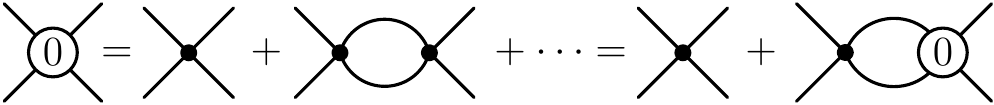}
\caption{Diagrammatic version of the Lippmann-Schwinger equation 
for the two-nucleon T-matrix at LO, depicted by the circled zero.  The solid 
lines represent nucleon fields, whereas the dot represents a contact 
interaction $C_{0,\sti}^{(0)}$.}
\label{fig:LO-T-Matrix}
\end{figure}

The LO calculation can be carried out in closed form by solving the 
Lippmann-Schwinger equation for a separable potentials $V_{2,\sti}^{(0)} = 
C_{0,\sti}^{(0)}\ket{g}\bra{g}$, where $\braket{p}{g} = g(p^2) = 
\exp({-}p^2/\Lambda^2)$ with a cutoff scale $\Lambda$ is a Gaussian regulator 
and $p$ is the $N\!N$ center-of-mass momentum.
This is shown diagrammatically in Fig.~\ref{fig:LO-T-Matrix}.
Some results discussed in Sec.~\ref{sec:Results} are obtained using a slightly 
different implementation, employing so-called ``dibaryon'' fields to describe the 
two-body sector and using a sharp momentum cutoff, which is essentially 
equivalent to choosing $\braket{p}{g}$ to be a step function.
This approach, deriving two- and three-body equations directly from Feynman
diagrams has been discussed in detail in 
Refs.~\cite{Konig:2015aka,Konig:2016iny}, so that here only the potential 
formalism with Gaussian regulator is considered.

Starting from $t_\sti^{(0)} = V_{2,\sti}^{(0)} + V_{2,\sti}^{(0)} G_0 
t_\sti^{(0)}$, where $G_0$ is the free two-body Green's function, the separable 
form of the interaction makes it possible to directly write down the solution as
\begin{equation}
 t_\sti^{(0)}(z;\veck,\veck')
 = \mbraket{\veck}{t_\sti^{(0)}}{\veck'} = g(k^2) \tau_\sti(z) g(k'^2)
 \mathtext{,}
 \tau_\sti(z) = \left[1/{C_{0,\sti}^{(0)}} - \mbraket{g}{G_0}{g}\right]^{-1}
 \,,
\label{eq:T2-LO}
\end{equation}
where $z$ denotes the energy.
$C_{0,\sti}^{(0)}$ can now be determined by matching this T-matrix to the
effective range expansion at the on-shell point, $\veck=\veck'$ and $E = 
k^2/\MN$.
The unitarity limit (infinite scattering length, $1/a_\sti=0$) is reproduced by
setting $C_{0,\sti}^{(0)} = \frac{{-}2\pi^2}{\MN\Lambda}\theta^{-1}$,
where $\theta = 1/\sqrt{2\pi}$ for the Gaussian regulator used here.
This means that the $C_{0,\sti}^{(0)}$ do not introduce any physical parameters.
At NLO, the correction to the T-matrix is
\begin{equation}
 t_\sti^{(1)} = V_{2,\sti}^{(1)}
 + V_{2,\sti}^{(1)} G_0 t_\sti^{(0)} + t_\sti^{(0)} G_0 V_{2,\sti}^{(1)}
 + t_\sti^{(0)} G_0 V_{2,\sti}^{(1)} G_0 t_\sti^{(0)} \,,
\label{eq:T2-NLO}
\end{equation}
\ie, the sum of all possible terms linear in $V_{2,\sti}^{(1)}$.
The overall NLO T-matrices $t_\sti^{(0)} + t_\sti^{(1)}$ should reproduce the
physical values of the $N\!N$ S-wave scattering lengths, which leads to 
$C_{0,\sti}^{(1)} = \frac{\MN}{4\pi a_\sti} C_{0,\sti}^{(0)2}$.
Note that instead of using Eq.~\eqref{eq:T2-NLO} it is possible to conveniently
obtain $t_\sti^{(1)}$ as well as higher-order corrections by solving integral 
equations similar to the one defining $t_\sti^{(0)}$.
Details about this procedure can be found in 
Refs.~\cite{Vanasse:2013sda,Konig:2016iny}.

With the two-body LECs determined, it is possible to proceed with calculations 
for three and four nucleons.
In the following the unified Faddeev + Faddeev-Yakubovsky (F+FY)
framework that was used to obtain the four-nucleon results presented in
Ref.~\cite{Konig:2016utl}.
The approach follows 
Refs.~\cite{Platter:2004he,Platter:2004zs,Platter:2005,Platter:2005sj} (which, 
in turn, are based on the work of Kamada and Glöckle~\cite{Kamada:1992aa}) but 
uses an independently developed numerical implementation.
Since a fully comprehensive description of the method is beyond the scope of this
work, emphasis is put here primarily on details regarding the perturbative 
treatment of NLO contributions.

\paragraph{Three nucleons}

It is a distinct feature of pionless EFT that a three-nucleon interaction 
(3NI) enters at LO in the power counting, while naïvely it would be expected to 
contribute only much later.
This promotion of the 3NI is a direct consequence of the unnaturally large $N\!N$
S-wave scattering length, leading to the triton as an effective Efimov
state~\cite{Bedaque:1998kg,Bedaque:1998km,Bedaque:1999ve}.
In the separable potential formalism the LO 3NI can be implemented as
\begin{equation}
 V_3^{(0)} = D_0^{(0)} \, \ket{\Triton}\ket{\xi}\bra{\xi}\bra{\Triton} \,,
\label{eq:V3-0}
\end{equation}
where $\ket{\Triton}$ projects onto a $J=T=\nicefrac12$ three-nucleon state and 
the regulator is now defined, for Jacobi momenta $\vecu_1 = 
\frac12(\veck_1-\veck_2)$ and $\vecu_2 = 
\frac23[\veck_3-\frac12(\veck_1+\veck_2)]$, as $\braket{\vecu_1 \vecu_2}{\xi} 
= g\big(u_1^2+\frac{3}{4}u_2^2\big)$.
The  $\veck_i$ label the individual nucleon momenta.
The NLO correction $V_3^{(1)}$ has the same form as Eq.~\eqref{eq:V3-0}, but 
involves the LEC $D_0^{(1)}$.

The Faddeev equations in an abstract operator notation take the form
\begin{subalign}[eq:Faddeev]
 \ket{\psi} &= G_0\,t^{(0)}\,P \ket{\psi}
  + G_0\,t^{(0)}\,\ket{\psi_3} \,,
 \label{eq:Faddeev-a} \\
 \ket{\psi_3} &= G_0\,t_3^{(0)}\,(1+P) \ket{\psi} \,,
 \label{eq:Faddeev-b}
\end{subalign}
where $\ket{\psi} = \ket{\psi_{(12)3}}$ is one of the (equivalent) two-body
Faddeev components and $\ket{\psi_3}$ is defined in terms of the three-body
interaction $V_3$~\cite{Platter:2005}.
$G_0$ now denotes the free three-body Green's function and $P = P_{12}P_{23} +
P_{13}P_{23}$ generates the non-explicit components through permutations.
$t^{(0)}$ collectively denotes the two-body T-matrices $t_\sti^{(0)}$, whereas 
$t_3^{(0)}$ is the solution of Lippmann-Schwinger like equation with 
$V_3^{(0)}$ as driving term.
The equations are solved by projecting onto momentum states $\ket{u_1u_2;s}$,
where $s = \ket{%
 \couple{\ell_2}{\couple{\couple{\ell_1}{s_1}{j_1}}{\tfrac12}{s_2}}{J};
 \couple{t_1}{\tfrac12}{T}
}$ collects the relevant angular momentum, spin, and isospin quantum numbers,
coupled such that $\couple{\ell_1}{s_1}{j_1}$ describes the two-nucleon 
subsystem and $\ell_2$ denotes the orbital angular momentum associated with the 
Jacobi momentum $u_2$.
Since only S-wave interactions enter to the order considered here, all sums
over $s$ are naturally truncated to involve only states with
$\couple{\ell_1}{s_1}{j_1} = \sti = \OneSNot,\ThreeSOne$.
For details regarding the implementation and solution of 
Eqs.~\eqref{eq:Faddeev}, see 
Refs.~\cite{Gloeckle:1983,Stadler:1991zz,Platter:2005}, noting
that the coupling scheme used here for $\ket{s}$ is a somewhat unusual choice
for $3N$ calculations; it is chosen in order to be consistent with the 
four-nucleon states introduced below.

In order to calculate the NLO triton energy, the full LO wavefunction 
$\ket{\Psi} = (1 + P) \ket{\psi} + \ket{\psi_3}$ is required.
Assuming it to be normalized such that $\braket{\Psi}{\Psi}=1$, the NLO energy 
shift is given by
\begin{equation}
 \Delta E = \mbraket{\Psi}{V_\NLO}{\Psi}
 \mathtext{,} V_\NLO = \sum\nolimits_\sti V_{2,\sti}^{(1)} + V_3^{(1)} \,.
\end{equation}
To check the calculation, one can verify that ${\mbraket{\Psi}{H_\LO}{\Psi}}$
with $H_\LO = H_0 + \sum_\sti V_{2\sti}^{(0)} + V_3^{(0)}$ gives the same energy 
as obtained directly from Eqs.~\eqref{eq:Faddeev}.

While for the Faddeev equations only the potential between a single pair of 
nucleons (chosen to be nucleons 1 and 2) is needed explicitly, evaluating 
matrix elements requires the \emph{full} two-body potential including all 
pairwise interactions.
Temporarily dropping sub- and superscripts for simplicity, this can be written
as~\cite{Stadler:1991zz}
\begin{equation}
 V_2 = V_{12}
 + (P_{12} P_{23}) V_{12} (P_{23} P_{12})
 + (P_{13} P_{23}) V_{12} (P_{23} P_{13}) \,.
\label{eq:V-2}
\end{equation}
Using the antisymmetry of the full wavefunction, $P_{ij} \ket{\Psi} = 
{-}\ket{\Psi} \;\forall\; P_{ij}$, one can write $V_2 \ket{\Psi} = (1 + P) 
V_{12} \ket{\Psi}$, and noting furthermore that $(1+P)^\dagger(1+P) = 3(1+P)$ 
gives $\mbraket{\Psi}{V_2}{\Psi} = 3\mbraket{\Psi}{V_{12}}{\Psi}$.
Similar simplifications together with $(1+P)\ket{\psi_3} = 3\ket{\psi_3}$ can be
applied to the norm and matrix elements of $H_0$.

Equations.~\eqref{eq:Faddeev} are solved to tune $D_0^{(0)}$ at LO (with the 
two-body S-waves at unitarity) such that the triton bound state comes out at 
its physical energy.
At NLO, where the finite physical scattering lengths are included \via the
$V_{2,\sti}^{(1)}$, there is a corresponding shift in the triton energy.
The LEC $D_0^{(1)}$ is adjusted such that this shift is compensated by
$V_3^{(1)}$, thus keeping the triton at its physical position.
Once this is done, all ingredients are in place to make predictions for 
four nucleons.

\paragraph{Four nucleons}

For the four-nucleon system, there are two distinct Faddeev-Yakubovsky 
components, $\ket{\psi_A}$ and $\ket{\psi_B}$, corresponding two 3+1 and 2+2 
cluster configurations of the four-body system.
For each of these components there is a natural set of Jacobi coordinates,
$(\vecu_1,\vecu_2,\vecu_3)$ and $(\vecv_1,\vecv_2,\vecv_3)$, respectively, of
which the former is a direct extension of the three-body Jacobi coordinates
(defining $\vecu_3$ as the relative momentum of the fourth particle with respect
to the center of mass of the other three).
For the 2+2 setup, $\vecv_1=\vecu_1$, $\vecv_3$ denotes the relative momentum 
in the (34) system, and $\vecv_2$ is defined at the relative momentum between 
the (12) and (34) subsystems.
Using the formalism of Refs.~\cite{Platter:2005,Kamada:1992aa}, the
Faddeev-Yakubovsky equations are written as
\begin{subalign}[eq:FaddeevYakubovsky]
 \ket{\psi_A} &= G_0 t^{(0)} P
 \big[(1 - P_{34})\ket{\psi_A} + \ket{\psi_B}\big]
 + \frac13(1 + G_0 t^{(0)}) G_0 V_3^{(0)} \ket{\Psi}
 \\
 \ket{\psi_B} &= G_0 t^{(0)} \tilde{P}
 \big[(1 - P_{34})\ket{\psi_A} + \ket{\psi_B}\big] \,,
\end{subalign}
where $\ket{\Psi} = (1 - P_{34} - P P_{34})(1 + P)\ket{\psi_A}
+ (1 + P)(1 + \tilde{P})\ket{\psi_B}$ is the full four-body wavefunction and 
$G_0$ now represents the free four-body Green's function.
In addition to the permutation operators already encountered in the three-body
system, Eqs.~\eqref{eq:FaddeevYakubovsky} involve the operators $P_{34}$ and
$\tilde{P} = P_{13}P_{24}$.

As discussed above for the three-body system, the FY equations are solved
in a partial-wave momentum basis, involving now two sets of Jacobi momenta 
defined and sums over channel states,
\begin{subalign}[eqs:Coupling-4]
 \ket{a} &= \ket{
   \couple{\ell_2}{\couple{\couple{\ell_1}{s_1}{j_1}}{\tfrac12}{s_2}}{j_2},
   \couple{\ell_3}{\tfrac12}{j_3},
   \couple{j_2}{j_3}{J};
   \couple{\couple{t_1}{\tfrac12}{t_2}}{\tfrac12}{T}
 } \,, 
 \label{eq:Coupling-4-u}
 \\
 \ket{b} &= \ket{ 
  \couple{\lambda_2}{\couple{\lambda_1}{\sigma_1}{\iota_1}}{\iota_2},
  \couple{\lambda_3}{\sigma_3}{\iota_3},
  \couple{\iota_2}{\iota_3}{J};
  \couple{\tau_2}{\tau_3}{T}
 } \,,
 \label{eq:Coupling-4-v}
\end{subalign}
which refer, respectively, to the 3+1 and 2+2 cluster setups.
The $\ket{a}$ are a natural extension of three-nucleon states $\ket{s}$, 
including the angular momentum $\ell_3$ associated with $\vecu_3$ as well as
spin and isospin $\tfrac12$ for the fourth nucleon into the overall coupling 
scheme.
For the $b$ states, $(\lambda_1,\sigma_1,\tau_1)$ and
$(\lambda_3,\sigma_3,\tau_3)$ are quantum numbers for the $(12)$ and $(34)$
two-body subsystems, respectively, where $\lambda_{1,3}$ are the angular momenta
associated with the Jacobi momenta $v_{1,3}$.
The separation between the clusters is described by the momentum $v_2$ and its
associated angular momentum $\lambda_2$.
The projection of Eqs.~\eqref{eq:FaddeevYakubovsky} yields a set of coupled
equations which, unlike the Faddeev equations, does not naturally truncate even
if all interactions are pure S-wave.
As a consequence it is necessary to truncate the sums in 
Eqs.~\eqref{eqs:Coupling-4} (\eg, by choosing all total angular
momenta $j_i$ and $\iota_i$ less than some $j_{\text{max}}$) and study the 
numerical convergence of results as $j_{\text{max}}$ is increased.
More details can be found in Ref.~\cite{Kamada:1992aa}.

\section{Results and discussion}
\label{sec:Results}

The convergence pattern of the unitarity expansion for the binding energies of 
light nuclei is summarized in Table~\ref{tab:Results}.
The deuteron remains a zero-energy bound state at \NLO and only moves to $1/(\MN
a_t^2)$ at \NNLO, see Ref.~\cite{Konig:2016iny} for an explicit 
calculation.  This is the case for both a pure expansion in $1/a_{t}$ 
(neglecting range correction) as well as for the paired unitarity expansion that 
includes effective ranges together with finite-$a$ corrections.  The dominant 
source of uncertainty for the deuteron energy comes from the $1/(Q_2 a_t)$ 
expansion, which still amounts to a 50\% effect at \NNLO.  Conservatively 
taking the experimental binding energy as reference for the uncertainty 
estimate gives $B_d^\NNLO = 1.41 \pm 1.12~\MeV$.

At each order the triton binding energy remains fixed at its physical value 
because it is used as input to tune the 3NI.
At LO, \ThreeH and \ThreeHe are degenerate by construction, but the splitting
between the two iso-doublet states is a prediction at \NLO.
As discussed in Ref.~\cite{Konig:2015aka}, range corrections cancel at this 
order because LO is isospin-symmetric.
The dominant effects that determine the splitting are thus electromagnetic 
corrections as well as the difference between the $np$ and $pp$ 
(Coulomb-modified) scattering lengths.
The unitarity expansion predicts the triton-helion energy splitting 
as $(0.92 \pm 0.18)~\MeV$ at \NLO, in good agreement with the 
experimental value $0.764~\MeV$.
At \NNLO the mixing between electromagnetic and range corrections introduces a 
divergence that requires an isospin-breaking 3NI to be promoted to this 
order~\cite{Konig:2016iny}.
For the unitarity expansion this means that a new input is required, which is 
most conveniently chosen to be the \ThreeHe binding energy.
Neglecting range corrections, however, Ref.~\cite{Konig:2016iny} finds good 
convergence up to \NNLO for an expansion that only includes finite scattering 
lengths and electromagnetic corrections in perturbation theory.

\begin{figure}[tb]
\centering
\includegraphics[width=0.68\textwidth,clip]{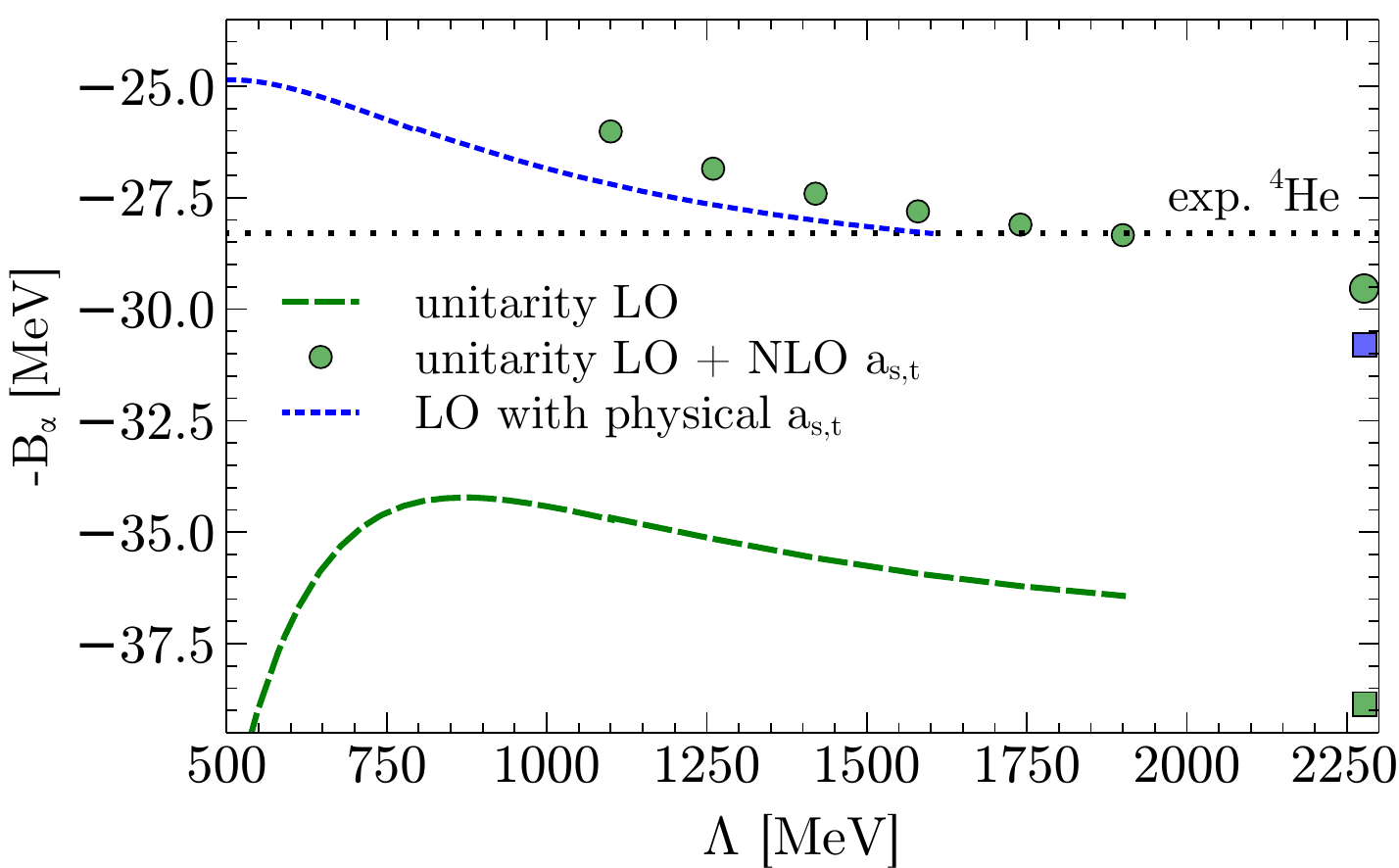}
\caption{%
\FourHe binding energy as function of the Gaussian cutoff.
(Blue) dotted and (green) dashed lines: standard Pionless EFT and full 
unitarity at LO, respectively.
(Green) circles first-order corrections in $1/a_{s,t}$ added in perturbation 
theory.
Large symbols on right edge: $\Lambda \to \infty$ extrapolation (see text).
}
\label{fig:En-4He-UU-atad}
\end{figure}

In the unitarity limit, \FourHe is formally equivalent to a system of four 
bosons.
It is known that each three-boson Efimov state with binding energy 
$B_3$ is associated with two four-boson states (tetramers)~\cite{Hammer:2006ct} 
at energies $B_4/B_3\simeq 4.611$ and $B_{4^*}/B_3\simeq 
1.002$~\cite{Deltuva:2010xd}.
The experimental values for the \FourHe 
ground and first excited states are, respectively, $B_\alpha/B_H\simeq 3.66$ 
and $B_{\alpha^*}/B_H\simeq 1.05$, where the \ThreeHe binding energy $B_H\simeq 
7.72~\MeV$ is used as reference to approximately account for electromagnetic 
corrections.
The closeness of these values to what is found in the unitarity 
limit suggests that a perturbative expansion can be expected to work well.
The numerical results shown in Fig.~\ref{fig:En-4He-UU-atad} confirm this
expectation.
The \FourHe binding energy as a function of the momentum cutoff $\Lambda$ is
found to converge as $\Lambda$ increases, indicating that the EFT calculation
is properly renormalized.
While any $\Lambda$ above the breakdown scale (of order $\Mpi$) is a valid 
choice in principle, quadratic polynomials in $1/\Lambda$ are fitted at large 
$\Lambda$ to quantitatively assess the convergence and conveniently extrapolate 
$\Lambda\to \infty$.
Figure~\ref{fig:En-4He-UU-atad} also shows a standard pionless 
calculation that includes finite $a_{s,t}$ at LO and gives results
consistent with Refs.~\cite{Platter:2004he,Platter:2004zs,Platter:2005}.
In the unitarity limit $B_\alpha = 39(12)~\MeV$ is found for the \FourHe
ground state.
In addition, there is a bound excited state just below the proton-triton 
breakup threshold.
Both these states are in excellent agreement with the universal unitarity 
expectation.

\begin{figure}[tb]
\centering
\includegraphics[width=0.68\textwidth,clip]{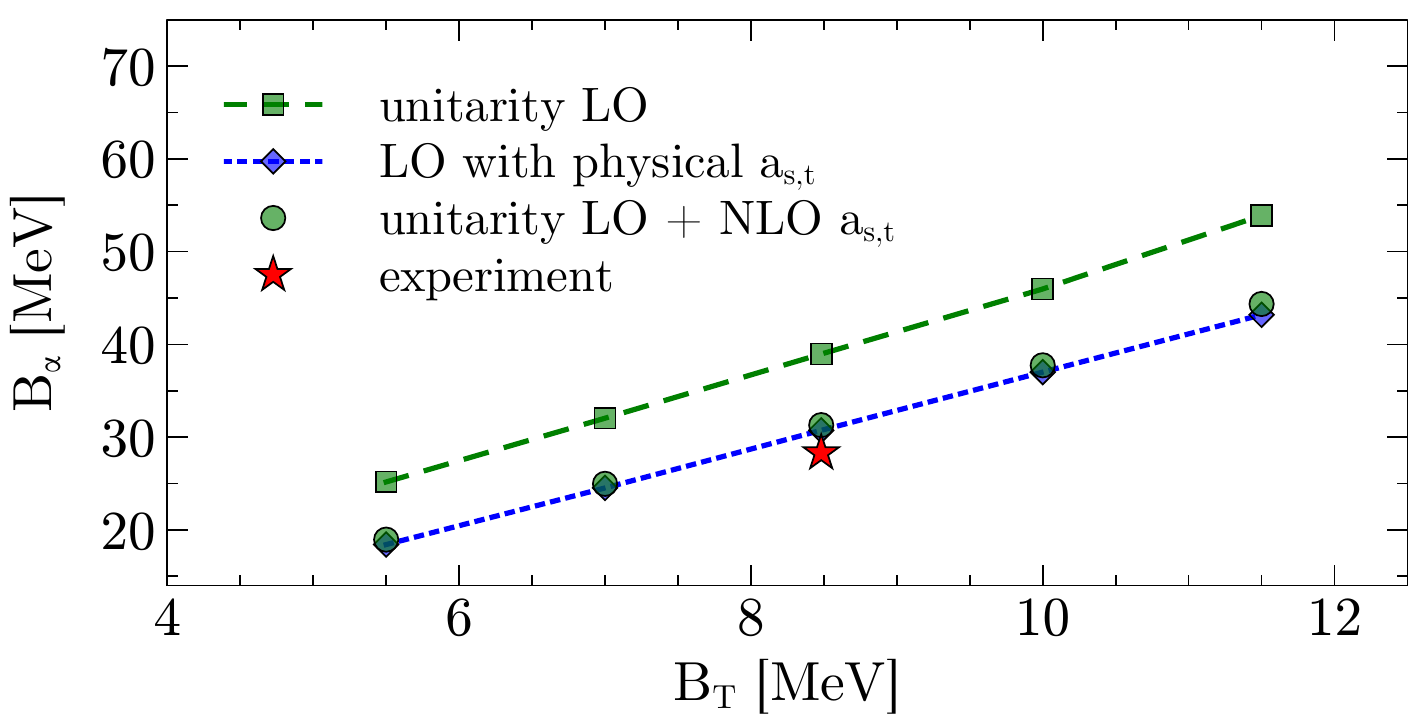}
\caption{%
Tjon line: correlation between the \FourHe and \ThreeH binding energies. 
(Blue) dotted curve: standard pionless LO result; (green) dashed upper curve: 
unitarity limit at LO.  Additional points nearly on top of the blue curve: 
inverse scattering lengths added in first-order perturbation theory.
Star: experimental point.
}
\label{fig:Tjon-UU}
\end{figure}

An incomplete \NLO (neglecting effective ranges and electromagnetic 
contributions) is calculated here to study the effect of finite-scattering 
length corrections in \FourHe.
The result, $30(9)~\MeV$ for $\Lambda\to\infty$ comes out 
very close the standard pionless LO calculation, indicating that the
$1/(Q_4a_{s,t})$ expansion works remarkably well up to this order.
The uncertainty of this value, as well as that of the LO result quoted above,
is $\OO(r_{s,t}/a_{s,t}) \simeq 30\%$ based on the expectation that range 
corrections are dominant in this case.
Importantly, $(B_\alpha/B_T)^{\NLO (r=0)}\approx3.48$ is also in good agreement 
with $(B_\alpha/B_T)^\text{exp}=3.34$.
As shown in Fig.~\ref{fig:Tjon-UU}, the rapid convergence persists off the 
physical point: the correlation between $3N$ and $4N$ binding energies 
(Tjon line) is perturbatively close to the unitarity result over a significant
range of energies.
While a proper calculation of the excited state is computationally very 
demanding due to a slow convergence of the FY calculation for a state so close
to a threshold, four-boson calculations performed using nuclear scales indicate 
that the $1/(Q_4a_{s,t})$ corrections furthermore push the bound excited state
into the continuum by about the amount expected from 
experiment~\cite{Konig:2016utl}.

Very recently it was found that a four-body forces is required to renormalize
the universal four-boson system once range corrections are included at 
\NLO~\cite{Bazak:2018qnu}.
This result directly carries over to pionless EFT---and thus to the unitarity 
expansion considered here---and implies that a new observable, most obviously 
taken to be the \FourHe binding energy, is required at this order to set the
scale of the four-body force.
Even with this additional required input the theory however remains predictive
for other four-body observables like \FourHe charge radius and excited 
state energy, as well as for heavier systems, assuming the unitarity expansion 
converges for these.

\begin{table}
\centering
\begin{tabular}{c|c|c|c||c}
\hline\hline
\rule{0pt}{1.2em} state
& $E_B^\LO/\MeV$ & $E_B^\NLO/\MeV$ & $E_B^\NNLO/\MeV$ & $E_B^{\text{exp.}}/\MeV$ 
\\
\hline
\rule{0pt}{1.2em}
\TwoH
\rule{0pt}{1.2em}
& $0$ & $0$ & $1.41 \pm 1.12$ & $2.22$\\
\ThreeH
& \underline{$8.48$} & \underline{$8.48$} & \underline{$8.48$} & $8.48$ \\
\ThreeHe
& $8.5 \pm 2.5$ & $7.6 \pm 0.2$ & \underline{$7.72$} & $7.72$ \\
\FourHe
& $39 \pm 12$ & $\;30 \pm 9^*$ & & $28.3$ \\
\hline\hline
\end{tabular}
\label{tab:Results}
\vspace{0.5em}
\caption{
Unitarity expansion convergence pattern.
Underlined values indicate energies which are used as input values to determine
three-body LECs.
An asterisk superscript indicates an incomplete \NLO result which only includes 
the finite-scattering length but no contributions from effective ranges or 
electromagnetic interactions.}
\end{table}


The unitarity expansion constitutes a paradigm shift in the EFT-based 
description of light nuclei, deemphasizing the importance of two-body details 
in favor of using the three-body sector as ``anchor point.''
As such, it is
not unlike more phenomenological approaches using input from heavier nuclei
in order to constrain few-nucleon forces.
It is, however, much more systematic by focusing on light nuclei and strives
to answer the question what really is essential to describe these systems.
As discussed compellingly in Ref.~\cite{Kolck:2017zzf}, the idea can be boiled 
down to interpreting discrete scale invariance, the most striking manifestation 
of which is the Efimov effect, as a fundamental principle governing nuclear 
physics.
In the bigger picture of things, the unitarity expansion furthermore stands in 
line with other recent results that suggest a fascinating simplification of 
nuclear physics.
For examples, it has been observed that the isotopic chain of helium can be
remarkably well described using a single-parameter model~\cite{Fossez:2018gae},
and more recently a correlation analogous to the Phillips line has been observed
between the $d$-$\alpha$ scattering length and the \SixLi binding 
energy~\cite{Lei:2018toi}

It is an exciting question how well
the unitarity expansion works beyond what has been calculated so far.
The observation that bosonic systems at unitarity exhibit
saturation for large numbers of particles~\cite{Carlson:2017txq} and recent
calculations of nuclear matter using interactions guided by 
unitarity~\cite{Kievsky:2018xsl} provide reasons to be optimistic.
However, it remains to be seen to what extent lessons from universal bosonic 
systems carry over to nucleons, where beyond the four-body sector the influence 
of Fermi statistics is expected to become important.
Concrete work looking at systems heavier than \FourHe as well as observables
beyond binding energies is currently in progress.

\medskip
\subsubsection{Acknowledgments.}
I would like to thank Harald Grießhammer, Hans-Werner Hammer, and Bira van 
Kolck for their collaboration as well as for many insightful discussions and 
comments on this manuscript.
This work was supported in part by the Deutsche Forschungsgemeinschaft (DFG, 
German Research Foundation) -- Projektnummer 279384907 -- SFB 1245 and by the 
ERC Grant No.~307986 STRONGINT.

\end{document}